\documentclass[a4paper,12pt]{article}

\usepackage[english]{babel}
\usepackage[T1]{fontenc}
\usepackage[latin1]{inputenc}
\usepackage{graphics}
\usepackage{graphicx}           
\usepackage{dcolumn}            
\usepackage{bm}                 
\usepackage{enumerate}

\begin{document}

\title{Coherence theory and coherence phenomena in a closed spin-1/2 system}
\author{Olavi Dannenberg\footnote{Author E-mail: {\sf olavi.dannenberg@helsinki.fi}} \\ {\small Helsinki Institute of Physics, PL 64, FIN--00014 Helsingin yliopisto, Finland}}
\date{May 15, 2008}

\maketitle


\begin{abstract}
A simplified Heisenberg spin model is studied in order to examine the idea of decoherence in closed quantum systems. For this purpose, we present a quantifiable definition to quantum coherence $\Xi$, and discuss in some detail a general coherence theory and its elementary results. As expected, decoherence is understood as a statistical process that is caused by the dynamics of the system, similar to the growth of entropy. It appears that coherence is an important measure that helps to understand quantum properties of a system, e.g., the decoherence time can be derived from the coherence function $\Xi(t)$, but not from the entropy dynamics. Moreover, the concept of decoherence time is applicable in closed and finite systems. However, in most cases, the decay of off-diagonal elements differs from the usual $\exp(-t/\tau_{\rm d})$ behaviour. For concreteness, we report the form of decoherence time $\tau_{\rm d}$ in a finite Heisenberg model with respect to the number of particles $N$, density $n_{\rho}$, spatial dimension $D$ and $\epsilon$ in a $\eta/r^{\epsilon}$-type of potential. 
\end{abstract}

\newcommand{\work}{{\fbox{\tiny work}}}
\newcommand{\ok}{{\fbox{\tiny ok}}}
\newcommand{\refok}{}                    
\newcommand{\dataref}{{$\bullet$}}                            
\hyphenation{Dannen-berg}
\hyphenation{Natur-wissen-schaften}

\section{Introduction}

Decoherence is widely accepted as an explanation of how quantum correlations are damped out to make physical systems effectively classical, and thus it provides a satisfactory answer to the dilemma of Schr\"odinger's poor cat \cite{sc35}. Open quantum systems with an infinite environment have been studied in great detail in various environment-induced decoherence schemes \cite{cl83, UZ89, zu91, hpz92, alzp95, dm2006} using master equations of reduced density matrices. However, the ``cosmological'' aspect of physical reality motivates studies in closed quantum systems since, strictly speaking, the universe has no environment \cite{UZ89, zu91, schlosshauer2004, sc2005} and yet decoherence is observed in our universe \cite{br96, nemes}. The most interesting questions are: How coherence in the universe should be modelled and understood? How reasonable it is to talk about decoherence in closed systems? There are three main lines in studying closed quantum systems: decoherent histories approach \cite{dh92, gmh93, bh1996, bh99, halliwell}, self-induced decoherence scheme \cite{cl1997, lc1998, castagnino1999, lcb1999, cl2005} and density matrix formalism. Density matrix formalism has also been used to study open and finite quantum systems \cite{schlosshauer2004, sc2005, zu82, om94}.

Decoherent (or consistent) histories approach uses the frame of the many histories interpretation of quantum mechanics. The source of decoherence is projections along histories that produce effective coarse-graining, and thus decoherence. Intuitively, histories are analogous to paths in path integral formalism, and meaningful histories should obey consistency conditions. The criticism against decoherent histories approach mainly consists of the following arguments: (1) The notion of ``decoherence'' in decoherent histories approach is not equivalent to ``decoherence'' in density matrix formalism \cite{gmh93}. (2) Philosophically, decoherent histories approach seem to violate one of the main principles of quantum physics, i.e., {\it in principle, the wave function is the complete description of the state of the system}. Histories seem to be the complete description and the wave function incomplete \cite{goldstein1998}. (3) In decoherent histories approach, decoherence effects are dependent on the chosen set of histories \cite{dk96, k96}. The choice of the set of histories is arbitrary and results in contrary propositions \cite{k97}, i.e., physical phenomena that are predicted (or retrodicted) using the consistent histories approach depend on the chosen set of histories. In Ref.~\cite{ghirardi1999} a peculiar feature of decoherent histories is shown: in a universe in which decoherent histories approach is valid, the objective properties are not objective at all, but dependent on chosen set (or family) of histories. This is a version of the {\it preferred basis problem} \cite{schlosshauer2004, barrett2003}.

In the self-induced decoherence scheme, decoherence is created within the undivided system. There is no need for environment as the source of decoherence, unlike in the environment-induced decoherence (in open systems). Decoherence appears because quantum correlations of expectation value $\langle {\rm M} \rangle$ in continuous energy spectrum are proportional to $\int {\rm d}E \int {\rm d}E' {\rm M}(E,E'){\rm e}^{-i(E-E')t/\hbar}$  and, according to Riemann-Lebesque theorem \cite{rs1975}, it vanishes when $t\rightarrow \infty$ if the integrand is $L^1$ integrable, and thus results in a diagonalisation of expectation value. There are three crucial assumptions: (1) continuous energy spectrum (i.e., the system is infinite), (2) the coefficients in the energy eigenbasis are integrable functions of energy (for finite systems this condition naturally fails), and (3) the time scale is infinite \cite{sc2005}. As analysed in Ref. \cite{sc2005}, for finite systems in long (but finite) time the diagonalisation does not occur. However, the critique against self-induced decoherence scheme presented in Ref. \cite{sc2005} is based on open and finite system analysis that is fundamentally different from closed and infinite system, as demonstrated in Section \ref{cohdyn}. Still, the definition of coherence in self-induced decoherence scheme lacks generality since it is applicable only to infinite systems.

The density matrix formalism is a natural way to study quantum systems, so it should be useful also in decoherence studies in closed systems. Moreover, it does not have the same drawbacks as the two other main lines to study closed systems. However, the concept of {\it coherence} is not clear at all (neither in density matrix formalism nor in any other formalism). A well established coherence theory using density matrices should help clarify the concept of decoherence in closed systems. Coherence theory and conceptual problems of decoherence (especially in closed and finite systems) are interesting and wide research topics alone, but related to this study, the following elementary results are important.
\begin{enumerate}
\item There is a basis-vector-set-independent way to quantify coherence uniquely. 

\item The possibility of {\it recoherence} does not mean that there is no decoherence. Decoherence is {\it a dynamical phenomenon that reduces coherence}, and hence, {\it recoherence is a dynamical phenomenon that increases coherence}. All finite quantum systems may experience recoherence.

\item There are two different coherence types (similar to the different entropy types): {\it idealistic} and {\it realistic} coherence. 
\begin{itemize}
\item The idealistic coherence scheme can be applied only by those observers who do not interact with the universe, and who know the wave function of the universe and its time evolution. In a closed system, there is no idealistic decoherence, since the wave function of the universe always remains pure.
 
\item The realistic coherence scheme is the internal view of the universe calculated from the wave function of the universe. It describes the events as the real observers that are totally correlated with the universe perceive them. To acquire this realistic viewpoint, one should make an effective theory from the wave function of the universe, e.g., to use reduced density matrices. This is often referred to as {\it coarse-graining}.
\end{itemize}
A short analysis of the results is presented in Section \ref{general}. In this study, the word ``decoherence'' refers to the concept of realistic decoherence. Exceptions are mentioned.
\end{enumerate}
Idealistic decoherence is often referred to with the word ``decoherence'', since, in open and infinite system studies both types of decoherence behave similarly. The principal difference between idealistic and realistic decoherence can be seen only in closed systems.

The first main purpose of our study is to define coherence uniquely and to analyse carefully the consequences of coherence defined as
\begin{equation}
\Xi(\rho)=\frac{N}{N-1}\left(\lambda_{\rm max}-\frac{1}{N} \right),
\label{koherenssi}
\end{equation}
where $\lambda_{\rm max}$ is the maximum eigenvalue of the density matrix $\rho$, and $N$ is the dimension of the Hilbert space of the corresponding quantum system. The definition satisfies the requirements for coherence and is useful in, e.g., calculating decoherence times of quantum systems. However, it is not an extensive measure like entropy, but intensive and probability-like. Both entropy and coherence are important concepts in understanding different properties of a quantum system.

Second, we apply the above presented coherence function to study a Heisenberg spin model in order to clarify the idea of decoherence in closed systems. We focus on the single particle coherence, and thereby calculate the time evolution of the (single particle) coherence of the system. This coarse-graining is similar to the one used in Gibbs coarse-grained entropy calculations \cite{gi60, sk93}. It is demonstrated that $\Xi$ is a valid and useful measure for coherence and coherence dynamics of the simple closed and finite system under the study. Moreover, we determine how the decoherence time depends on the system variables (particle number $N$, density $n_{\rho}$, potential and dimension $D$ of the system). The behaviour of decoherence is calculated for $\eta/r^{\epsilon}$ potentials in various dimensions $D$. The results are interesting: the usual coherence decay profile $\exp(-t/\tau_{\rm d})$ occurs only in special cases. Generally the decay follows 
\begin{equation}
\Xi_{\rm re}(t)\sim\exp{\left(-(t/\tau_{\rm d})^{f(D,\epsilon)}\right)},
\end{equation}
where $f(D,\epsilon)=1.97\left(1-0.93\exp{(-0.65D^{1.35}\epsilon^{-1.68})}\right)$, and
\begin{eqnarray}
\tau_{\rm d} \!\! & \!\!=\!\! & \!\! \eta^{-1}n_{\rho}^{-\epsilon/D}\left(\frac{N}{200}\right)^{-0.085(D-1)/\epsilon} \\
\!\!& \!\!\times\!\! & \!\!\!\! \left(0.29{\rm e}^{-0.79(D-1)\epsilon^{1/4} -0.13(\epsilon-3.4)^2} \!\!+\! 0.17\frac{\epsilon \!-\! 1}{2^\epsilon}((D\!-\!2)^2)^{0.19(\epsilon-1)} \!\!+\!\frac{0.03}{\epsilon^{1/2}} \!+\! 0.07\right) \nonumber
\end{eqnarray}
is the decoherence time.

This paper consists of six main sections. First, we review a relevant part of coherence theory in Section \ref{general}. The spin model is introduced in Section \ref{model}. In Section \ref{theory} we present theoretically the dynamics of a simple initial state, and apply the coherence function $\Xi(t)$ as the measure of the realistic coherence of the whole system. The structure of numerical simulations is presented in Section \ref{simu}, and Section \ref{results} explains the results of the numerical simulations along with an analytical reality-check of basis independence of our model. Section \ref{discussion} is for discussion.

\section{Introduction to general coherence theory \label{general}}

In a scientific dictionary, coherence is defined as ``correlation between the phases of two or more waves, so that interference effects may be produced between them, or a correlation between the phases of part of a single wave'' and ``property of moving in unison'' \cite{mcgrawhill}. The amount of coherence of waves depends on how correlated they are; how close the wave lengths, phases and amplitudes are to each other. Thus, coherence can generally be defined as ``correlation of wave-like entities'' and it enables interference. Quantum physical objects are generally described using wave functions or state vectors, which are ``wave-like entities''. Thus, quantum coherence could be understood as ``a measure of the strength of quantum correlations''. Within the framework of classical correlations, quantum correlation phenomena cannot be explained. Evidently, entanglement and superposition states in general can produce observable non-classical correlations \cite{ze}. In the density matrix formalism, the fingerprints of quantum correlations are the off-diagonal elements of the density matrix. From now on, in our article the word ``coherence'' mostly refers to quantum coherence.

Let $\rho$ be a density matrix of a closed system in the basis set $A=\{\vert \alpha_i\rangle\},i=1 \dots N$. The basis vectors span a $N$-dimensional Hilbert space $H$. The equivalent description about the (closed) system can be given with an arbitrary basis set $B=\{\vert \beta_i\rangle\}$ if the basis set spans the same Hilbert space $H$. The quantum dynamics of the system does not depend on the chosen basis vectors. 

Let us now consider quantum correlations (coherence) in a basis set, say $A$. The word ``quantum coherence'' refers to the off-diagonal elements of the density matrix $\rho$, but the height and location of off-diagonal elements (and diagonal elements) depend on the chosen basis set. Therefore, the ``pointer basis'' has been invented: it is the basis set in which the quantum dynamics diagonalises the (reduced) density matrix \cite{UZ89, zu91, zu82, zu81}. If one wants to describe the whole closed quantum system without using tricks to make it effectively open, one cannot use the pointer basis since the unitary dynamics does not diagonalise the (irreduced) density matrix in any basis. 

However, the question ``How much coherence is there in a given density matrix of a particular closed system?'' is a relevant one, and the answer should not depend on the chosen basis set. In a closed system, the quantity called ``coherence'' of the total density matrix $\rho$ should be a conserved quantity because of the previous reasoning. According to the Noether's principle \cite{alav4}, the ``conservation law of coherence'' is a result of a particular symmetry of the system. The best candidate for that symmetry is the invariance of the density matrix under the change of the basis vector set. If one wishes to define ``coherence'', the definition should obey these constraints. 

The density matrix of a totally coherent system of $N$ possible states described with, say basis set $A=\{\vert \alpha_i\rangle\},i=1 \dots N$,
\begin{equation}
\rho_A=\left(
\begin{array}{cccc}
\frac{1}{N} & \frac{1}{N} & \cdots & \frac{1}{N} \\
\frac{1}{N} & \frac{1}{N} & \cdots & \frac{1}{N} \\
\vdots & \vdots & \ddots & \vdots \\
\frac{1}{N} & \frac{1}{N} & \cdots & \frac{1}{N}
\end{array}
\right)
\end{equation}
has the description
\begin{equation}
\rho_B=\left(
\begin{array}{cccc}
1 & 0 & \cdots & 0 \\
0 & 0 & \cdots & 0 \\
\vdots & \vdots & \ddots & \vdots \\
0 & 0 & \cdots & 0
\end{array}
\right)
\end{equation}
in the eigenvalue basis $B=\{\vert \beta_i\rangle\}$ that is always diagonal. The transformation is obtained by unitary transformation $U$ so that $\rho_B=U\rho_A U^{-1}$. By previous reasoning, these density matrices should contain the same amount of coherence. And, because quantum correlations refer to correlations that cannot be explained by classical correlations, they should be measurable -- at least in the idealistic sense. The totally classical density matrix
\begin{equation}
\rho_C=\left(
\begin{array}{cccc}
\frac{1}{N} & 0 & \cdots & 0 \\
0 & \ddots & \cdots & \vdots \\
\vdots & \vdots & \ddots & \vdots \\
0 & 0 & \cdots & \frac{1}{N}
\end{array}
\right)
\end{equation}
is diagonal in any basis. Let us consider following measurement scheme: we are free to measure an observable $O$ in any kind of basis. The probability to observe ``a matter of fact'' $\vert O \rangle$ is
\begin{equation}
{\rm Prob}(\rho,O)={\rm Tr}(\rho O)={\rm Tr}(\rho \vert O \rangle \langle O \vert).
\end{equation}
While measuring $\rho_A$, the probability ${\rm Prob}(\rho_A,O)$ is maximised when $\vert O \rangle = \frac{1}{\sqrt{N}}\left(1 \cdots 1 \right)^T_A$, and thus ${\rm Prob}(\rho_A,O) =\frac{1}{N} \sum_{i,j}^N \rho_{i,j} =1$. For the totally classical density matrix $\rho_C$, the probability for the same measurement outcome is ${\rm Prob}(\rho_C,O) =\frac{1}{N} \sum_{i,j}^N \rho_{i,j} =\frac{1}{N}$. Thus, maximum quantum correlations (coherence) are causing a $1-\frac{1}{N}$ increase in the probability of the measurement outcome. Naturally, the same probabilities of measurement outcomes happen in the eigenvalue basis, so that the measured state is $\vert O'\rangle=U \vert O\rangle=\left(1 \: 0 \cdots 0 \right)^T_B$. The eigenvalue basis is useful, since the maximum eigenvalue $\lambda_{\rm max}$ is the probability for a state (of the measurement outcome) to be in its most probable state, i.e., to demonstrate maximally the quantum nature of a particular system.

Thus, we define coherence $\Xi$ of the density matrix $\rho$ by
\begin{equation}
\Xi(\rho)=\frac{N}{N-1}\left(\lambda_{{\rm max}, \rho}-\frac{1}{N}\right),
\label{eq1}
\end{equation}
where $N$ is the dimension of the density matrix $\rho$. Subtraction of the term $1/N$, which represents the classical probability, is a background correction that sets the minimum coherence to zero. The prefactor $\frac{N}{N-1}$ normalises the maximum coherence to 1, since a maximally coherent system would yield a probability-optimised measurement outcome with probability of 1. The value of $\Xi$ describes how strongly the quantum coherence influences the behaviour of a particular system as whole. Since the value of coherence $\Xi$ depends only on the dimension of the Hilbert space ($N$) and the maximum eigenvalue $\lambda_{\rm max}$ of the density matrix $\rho$, it clearly is a basis-set-independent measure.

\subsection{About coherence dynamics \label{cohdyn}}

There are some important remarks concerning coherence dynamics of closed quantum systems. At first, let us consider a classical example of, say $10^9$, longcase pendulum clocks that do not interact with each other. They are similar to each other except that their periods are random, due to random effective lengths of pendulums. At the instant of time $t_0$ the phase of all pendulums is the same, i.e., the pendulums are at the lowest possible point and swinging leftwards. Classically thinking, this initial setting is very coherent in the phases of pendulums. After a short time, at the moment $t_1$ the phase coherence is totally lost. However, at the moment $t_{\rm P}$ the phase coherence has revived -- pendulums are at the lowest possible point and swinging leftwards -- since the number of pendulums is finite and they all have fixed periods. The revival time $t_{\rm P}$ is known as {\it Poincar\'e recurrence} time, and is the longest possible timescale in the system. In an infinite time, a finite system recurs arbitrarily close to its initial state (or any state) arbitrarily many times. \cite{gag1997}

Let us consider now a Gaussian wave packet that is trapped in an anharmonic potential $U(r)=\frac{1}{2}m\omega^2(r-r_0)^2+\chi (r-r_0)^3$ that has small anharmonicity $\chi$. At first, the wave packet oscillates coherently in the potential. After an intermediate time the wave packet has lost its form and shape, i.e., has lost its coherence. However, after a very long time the wave packet reforms again to Gaussian shape if the system consists of a finite amount of bound states. For more details, see Ref. \cite{gs1995} and references therein. The behaviour of coherence can be studied via the time evolution of the expectation value of position $\langle x(t) \rangle$. The strong oscillations of the expectation value of the position present in the beginning, dampen as the system evolves in time, resulting in a fluctuating mean.

The third interesting example is interacting bosons in a boson field so that the group of bosons forms the system and the boson field forms the environment. The standard procedure to study the problem is to trace out the boson field from the complete description density matrix in order to acquire the master equation for the density matrix of the system. The system is an open one. See, e.g., Refs. \cite{ms1991}: Ch. 14, \cite{wm1994}: Ch. 6.1, \cite{carmichael1999}: Ch. 1, \cite{bp2002}: Ch. 3.3. At first the boson field is treated to have a finite (but large) number of degrees of freedom, but when correlation functions should be calculated, the number of degrees of freedom of the boson field is assumed to be infinite (justified by the fact that it is very large compared to the number of degrees of freedom of the system), so that the sums can be converted to calculable integrals. These kinds of environments are accepted sources of decoherence, resulting in a monotonic decay of coherence that has characteristic decoherence time \cite{dm2006}.

In all these examples the concept ``coherence'' (classical or quantum) is reasonably applicable, but the concept ``decoherence'' is commonly associated only with the third example because that example describes a net loss of coherence out of the system. The fundamental difference between the systems of the first two examples and the last is that the system of the last example is an {\it open system} and an {\it infinite system}. The net loss of coherence requires both ``openness'' and ``infiniteness'' -- openness because an environment is needed for coherence dumping, and infiniteness because from a finite environment all coherence will eventually come back to the system. Thus, the observed net loss of coherence in some open quantum system studies is the result of an approximation that assumes infinity because of the need of calculating integrals instead of sums. It is an approximation that changes the fundamental behaviour of the system: it results in a monotonic decay of coherence with no revivals. Usually, in the fundamental level, approximations neither increase accuracy nor enlarge the area of application. Moreover, it is not even clear that our universe is infinite. The Bekenstein bound \cite{bekenstein1981} sets the upper limit to information (related to the degrees of freedom) in finite volume, and estimated (quantum) computational capacity of our universe is finite \cite{lloyd2002}.

The longcase pendulum clock example demonstrates the difference between a finite closed system and an infinite closed system: if one increases the amount of clocks, Poincar\'e recurrence time either increases or remains the same. There is no increase of recurrence time if the period of the added clock multiplied by an integer is the period of the existing clock system. If periods are random real numbers (it does not matter if the interval is fixed), for infinite amount of clocks the Poincar\'e recurrence time is infinite, i.e., there is no recurrence. Even if the system is closed, the decay of phase coherence is monotonic. If one considers the second example system with infinite setting, this feature is even more striking. The system is closed, so that there is no net loss of coherence out of the system. Infiniteness means there is no Poincar\'e recurrences, because the initially very narrowly concentrated finite amount of coherence spreads evenly to infinitely small amounts of correlations between all the infinite states -- and it never finds its way back. 

While closed and finite systems differ fundamentally from open and infinite systems, it appears that the term ``decoherence'' is an accurate description for what happened in these thought experiments. The word ``decoherence'' stands for a {\it dynamical phenomenon that reduces coherence}. The word ``decoherence'' does not even suggest that this reduction of coherence should be permanent. Finite systems may experience both decoherence and {\it recoherence}. Recoherence is a {\it dynamical phenomenon that increases coherence}. In density matrix formalism, decoherence is commonly understood as {\it the decay of off-diagonal elements of the density matrix}. More precisely, in general case, decoherence of a particular density matrix $\rho$ is the decay of the function $\Xi(\rho)$.

\subsection{Realistic and idealistic standpoints}

Similar kind of questions about the behaviour of closed systems have been a leading topic of debate over the entropy and universe more than hundred years. Here we will give only a brief introduction to this topic. A more interested reader is encouraged to study the vast literature covering the issue, starting with \cite{gi60, sk93, ee59, sk74, pe79, re95, br03} and references therein. 

Entropy is an ``extensive measure depending on the number of possible (micro)states and corresponding probabilities''. The probabilities of microstates are used as weights in Gibbs entropy $S_{\rm G}=-\sum_{i=1}^D p_i \ln p_i$.  Gibbsian entropy is often titled as ``statistical entropy'', as well as its quantum equivalent, the von Neumann entropy $S_{\rm N}=-{\rm Tr}\rho\ln\rho$. Entropy obeys the same ``independence of the basis vector set'' principle as coherence. 

The second ``law'' of thermodynamics states that the entropy of a closed system cannot decrease. However, the underlying laws of physics are reversible in time, and if the closed system is finite, then it should experience Poincar\'e recurrences, which means a decrease of entropy at some moments. Another entropy problem that has relevance to coherence studies was the observation that the entropy of a closed system (universe) studied as whole was a constant of motion while inside the universe it seems that entropy is increasing basically everywhere.

The irreversible entropy dynamics that is associated with the second ``law'' was a result of an approximation in deriving the Boltzmannian H-theorem \cite{sk74, pe79}. However, even Boltzmann himself noticed that irreversible entropy dynamics is not a lawlike theory but merely a statistical theory \cite{bo96} that allows the decrease of entropy. The second entropy problem was partially solved by introducing a method to {\it coarse-grain} the complete system description so that the statistical behaviour of entropy remained unchanged in closed and finite systems. In Gibbsian coarse-graining, probabilities of infinitely small phase-space elements are averaged in a finite phase-space element, and then entropies of all averaged probabilities are summed to obtain the entropy of a closed system \cite{gi60, sk93}. However, this did not end the philosophical discussion in which the dissatisfaction concentrated basically in two kind of arguments: ``what is the `right' coarse-graining to choose from the many possibilities (i.e., how to divide the system into subsystems)'' and ``while the `true' entropy of closed system is a constant, what does it mean that coarse-grained entropy changes in time''.

Let us now introduce two notations for entropy and coherence of a quantum system described with the density matrix $\rho$: {\it idealistic entropy} $S_{\rm id}=-\!{\rm Tr}\rho\ln\!\rho$, {\it idealistic coherence} $\Xi_{\rm id}\!=\!\Xi(\rho)$ and {\it realistic entropy} $S_{\rm re}\!=\!-\!\sum_i\!{\rm Tr}\rho_i \!\ln\!\rho_i$, {\it realistic coherence} $\Xi_{\rm re}=N^{-1}\sum^N_i\Xi(\rho_i)$, where $\rho_i$ is the density matrix of a subsystem. For any closed system, idealistic entropy and coherence are constants of motion. In calculating idealistic entropy and coherence, everything inside the closed system is treated quantum mechanically, and if the universe is under consideration, it still demands that measurements, measurement apparatuses and even observers are treated quantum mechanically. Measurements are expressed as interactions between measurement apparatuses and measured systems and so on. Clearly this is possible only for an idealistic observer, say, a Clever Chinchilla. She does not interact with the universe (closed system), and after having guessed the correct wave function and its time evolution, studies the universe only as an academic example in her brain (or with a pen and paper) without ever interacting with it.

The realistic standpoint describes what the real observers inside the universe observe. These observers are totally correlated with the universe and can only in their wildest imagination reach the Clever Chinchilla's view of the universe -- as we do in this article, when we refer to idealistic quantities. If the Clever Chinchilla wishes to study the universe of her academic example in order to figure out what the universe looks like when viewed from inside, she coarse-grains her perfect model and obtains realistic quantities. Naturally, idealistic and realistic quantities behave differently, but there is no contradiction between them -- if the coarse-graining is done right (note that the Clever Chinchilla is so clever that she never makes mistakes), she obtains as accurate a description as possible of the different possibilities of the universe as the Real Observer would see it. Because quantum processes are truly random, the Clever Chinchilla only acquires all possible measurement outcomes weighted with their probabilities. 

Both realistic and idealistic quantities describe the same system, but from different standpoints. Entropies are related via the quantity $I=S_{\rm re}-S_{\rm id}$, which is called {\it mutual information} \cite{alav2}. Mutual information expresses the amount of correlations between the (arbitrarily chosen) subsystems. Close relations between entropy and coherence suggest that there exists also a quantity $E=\Xi_{\rm id}-\Xi_{\rm re}$ that quantifies the degree of {\it mutual entanglement}, i.e., how entangled the subsystems are with each other. This quantification scheme for entanglement is applicable for all quantum systems in density matrix formalism.

The perfect description of the system is impossible inside the closed system, even in principle. The fundamental logical difference between idealistic and realistic standpoints is similar to G\"odel's incompleteness theorem \cite{go31}, which states roughly that in a richer logical system than the first order predicate logics (the universe) that has propositions concerning its own consistency (Real Observer, with the help of measurements, searches the perfect description of the universe) there exist propositions $\varphi_i$ of the form of ``$\varphi_i$ cannot be proved true''. To prove $\varphi_i$ requires a higher order metatheory (Clever Chinchilla). Thus, the closed system cannot be completely described within the system, and the complete and incomplete descriptions are different.

Decoherence is an {\it observed} phenomenon \cite{br96, nemes} in the universe. Thus, the decay of coherence should refer to the quantity $\Xi_{\rm re}$ -- even if the whole universe obeys unitary dynamics. There is, however, one problem without a proper answer: how to make coarse-graining correctly -- how to divide a closed system into subsystems. In this study, we coarse-grain to the smallest level of individual particles. A more detailed analysis of coherence theory and problems related to defining decoherence will be published elsewhere \cite{da03}.

\subsection{Entropy and coherence}

Both entropy and coherence are important concepts that help us to classify and understand properties
of quantum systems. Entropy is an extensive measure, while coherence is intensive and behaves like
probability \cite{entropiasta}. Let us illustrate some properties of entropy compared with properties
of coherence by considering two identical uncorrelated spin-$\frac{1}{2}$ systems that both have,
say, $\rho_{\frac{1}{2}}=\left(\begin{array}{cc}
\frac{1}{2} & \frac{1}{4} \\
\frac{1}{4} & \frac{1}{2}
\end{array}\right)$. The eigenvalues are $\lambda_{\pm}=\frac{1}{2}\pm \frac{1}{4}$. Thus, both
systems have coherence $\Xi_{\frac{1}{2}}=\frac{1}{2}$ and entropy $S_{\frac{1}{2}}\simeq 0.56$.
The system $\rho_{1+2}=\rho_1\otimes\rho_2$ has eigenvalues $\lambda= \frac{9}{16} \wedge
\frac{3}{16} \wedge \frac{3}{16} \wedge \frac{1}{16}$, and thus the entropy is $S_{1+2}\simeq
1.12$ and entropy per particle is $S_{\rm norm}\simeq 0.56$, and coherence is
$\Xi_{1+2}=\frac{5}{12}$. Moreover, constructing a greater system from uncorrelated smaller
systems causes a dramatical drop in coherence, since the coherence of $N$ joint uncorrelated
systems is proportional to $\prod_l^N \lambda_{{\rm max},l}$. 

However, the above mentioned ``anomaly'' in coherence is not surprising at all, if one remembers that coherence is closely related to probability, that is, to the maximum possible probability of the system. If one has the probability of $3/4$ to measure the maximum probability state of a two-level system, the probability of maximum probability state for two such systems is $(3/4)^2$. 

If one considers unitary dynamics, both (idealistic) coherence and (idealistic) entropy of closed system are constants of motion that give constraints to the system. In the system of $\rho_{1+2}$ the transition to the system that has eigenvalues $\frac{9}{16} \wedge \frac{7}{16}$ would be ``legal'' from coherence-point-of-view, but it would change the entropy $S_{1+2}\simeq 1.12 \rightarrow 0.69$, and while eigenvalues $\lambda\simeq 0.5 \wedge 0.32333 \wedge 0.125 \wedge 0.05167$ are possible from entropy-point-of-view, they would result in change of coherence of the system. Coherence has significant importance if one wishes to compare quantum properties of two systems that, e.g., have the same entropy. Of those, the system that has the greater coherence also has more quantum correlations, and would be more useful to, e.g., quantum computation. Entropy does not tell us the whole story.

\section{The model \label{model}}

We have decided to study the Heisenberg spin model because it is simple enough to solve, and yet complicated enough to simulate the properties of real quantum systems. Coupled spin systems are interesting from a quantum computational point of view, too. Our system, $N$ interacting particles fixed in a space, has no environment and, in that sense, the system forms a closed quantum universe. The particles are spin-$\frac{1}{2}$ particles, and the interaction between them is due to their spin-$z$ component (analogous to the Ising model). When there is no coupling with the environment (i.e., no outer environment), the two spin states have the same energy, which is taken to be zero. Zurek \cite{zu82}, Omn\`es \cite{om94} and Schlosshauer \cite{schlosshauer2004, sc2005} have considered a similar, but simpler model in order to study the decay of off-diagonal elements (i.e., quantum correlations) of a reduced density matrix. They labeled one particle as {\it the system}, and the others as {\it the environment} \cite{alav}, but, in this case we study the particle system as a whole.

The interaction Hamiltonian,
\begin{equation}
H=\hbar\sum^{N-1}_{j=1}\sum^{N}_{i=j+1}g_{ij}\sigma^{j}_{z}\otimes \sigma^{i}_{z}\prod^{N}_{k=1, \ne j,i}\otimes\mathbf{1}_{k},
\label{f0}
\end{equation}
describes the dynamics of the system. The interaction matrix $G$, where $g_{ij}=g_{ji}$, gives the interaction strength between particles $i$ and $j$. The interaction strength arises from the potential $V$; but, for formal calculations there is no need to know more about it, because particles are doomed to stay in one place. Fixing the positions of the particles is a justified assumption in decoherence studies since, in most cases, the decoherence time scale is the shortest time scale \cite{UZ89}, at least shorter than the time scale of particle motion. In numerical simulations, only potentials of the type $V=\eta/r^{\epsilon}$ are considered.

\section{Theoretical calculations \label{theory}}

Let us now consider a simple case in order to present our method. The task is to solve the time evolution of the realistic coherence $\Xi_{\rm re}(t)$ of a closed and finite system that is initially in a product state of superposition states
\begin{equation}
\vert \Psi(0)\rangle=\prod^{N}_{k=1}\otimes\left( a_{k}\vert+_{k}\rangle+b_{k}\vert-_{k}\rangle\right),
\label{f1}
\end{equation}
where $a$'s and $b$'s are normalised probability amplitudes $\vert a_{k}\vert^{2}+\vert b_{k}\vert^{2}=1$ for all $k=1,\dots,N$. The Schr\"odinger equation,
\begin{equation}
{\rm i}\hbar\partial_{t}\vert \Psi(t)\rangle=H\vert \Psi(t)\rangle,
\end{equation}
gives the dynamics of the system, and with the given initial condition of equation (\ref{f1}) one gets the time dependence
\begin{equation}
\vert \Psi(t)\rangle={\rm exp}\left[-{\rm i}\sum^{N-1}_{j=1}\sum^{N}_{i=j+1}g_{ij} \sigma^{j}_{z} \sigma^{i}_{z}t\right]\prod^{N}_{k=1}\otimes \left(a_{k}\vert+_{k} \rangle+b_{k}\vert-_{k}\rangle\right).
\end{equation}

The fate of the $l^{{\rm th}}$ particle is solved by tracing over other particles, i.e., degrees of freedom,
\begin{equation}
\rho_{l}={\rm Tr}_{1,\dots,N \ne l}\, \rho \,,
\end{equation}
where $\rho=\vert\Psi(t)\rangle\langle\Psi(t)\vert$. This is a crucial step. With Gibbsian coarse-graining \cite{gi60, sk93} we make an effective theory of our particle system by tracing over the ``uninteresting'' particles (that form an effective environment to the particular particle of our interest), as in the mean field approximation. The net effect of traced-out particles is described in a simpler form and with less degrees of freedom. 

We thus have
\begin{eqnarray}
\rho_{l} & = & \vert a_{l}\vert^{2}\vert+_{l}\rangle\langle+_{l}\vert+ \vert b_{l}\vert^{2} \vert-_{l}\rangle\langle-_{l}\vert \nonumber \\
& + & \left[a_{l}b^{*}_{l}\!\!\prod^{N}_{k=1,\ne l}\!\!\left( \vert a_{k}\vert^{2}{\rm e}^{-{\rm i} 2g_{lk}t}+\vert b_{k}\vert^{2}{\rm e}^{{\rm i} 2g_{lk}t}\right) \vert+_{l}\rangle\langle-_{l}\vert + {\rm h.~c.} \right].
\label{f2}
\end{eqnarray}
It is interesting that the result of Eq. (\ref{f2}) is the same as in Ref. \cite{zu82}, if one drops off the index $l$. In Ref. \cite{zu82} only an interactionless environment has been considered, but our model counts all the interactions between particles. Let us make our notation a bit lighter by denoting
\begin{equation}
z_{l}=a_{l}b^{*}_{l}\prod^{N}_{k=1, \ne l}\left(\vert a_{k}\vert^{2}{\rm e}^{-{\rm i} 2g_{lk}t}+\vert
b_{k}\vert^{2}{\rm e}^{{\rm i} 2g_{lk}t}\right).
\label{eqxyz}
\end{equation}
This $z_{l}$ (or its complex conjugate $z^{*}_{l}$) describes the fate of the off-diagonal elements of $l^{\rm th}$ particle. The eigenvalues of the reduced density matrix $\rho_l$ are
\begin{equation}
\lambda_{i,l}=\frac{1}{2}\pm\frac{1}{2}\sqrt{1-4(\vert a_l\vert^2 \vert b_l\vert^2-\vert z_l(t)\vert^2)}.
\label{eqeig}
\end{equation}

Now the time evolution of single particle coherences $\Xi_{l,{\rm single}}=\Xi(\rho_l)$ and the realistic coherence $\Xi_{\rm re}(t)=N^{-1} \sum_{l=1}^N \Xi_{l,{\rm single}}$ of the closed system can be evaluated. Inserting maximum eigenvalues of Eq. (\ref{eqeig}) into the definition of Eq. (\ref{eq1}) results in 
\begin{eqnarray}
\Xi_{\rm re}(t) \!\!\! & \!\!\!=\!\!\! & \!\!\! \frac{1}{N}\sum_{l=1}^N \frac{M}{M\!\!-\!\!1}\!\left(\!\lambda_{{\rm max},l}\!-\!\frac{1}{M}\right)\!
 = \! \frac{1}{N}\sum_{l=1}^N \! \frac{2}{1}\left(\frac{1}{2} \! + \! \frac{1}{2}\sqrt{1 \! - \! 4(\vert a_l\vert^2 \vert b_l\vert^2 \! - \! \vert z_l(t)\vert^2)} \! - \! \frac{1}{2}\right) \nonumber \\
\!\!\! & \!\!\!=\!\!\! & \!\!\! \frac{1}{N}\sum_{l=1}^N \sqrt{1-4(\vert a_l\vert^2 \vert b_l\vert^2-\vert z_l(t)\vert^2)},
\label{eq9}
\end{eqnarray}
where $N$ is the number of particles and $M$ is the dimension of the reduced density matrix.

\section{Numerical simulations \label{simu}}

The main aim of our numerical simulations is to demonstrate the usefulness of the definition of coherence presented in Section \ref{general}. The secondary aim is to study coherence behaviour of a simple closed and finite system, including Poincar\'e recurrences. The third aim is to calculate the dependence of {\it decoherence time} $\tau_{\rm d}$ of relevant system variables, i.e., $\tau_{\rm d}= \tau_{\rm d}(N, n_{\rho}, \eta, \epsilon, D)$, where $N$ is the number of particles, $n_{\rho}$ the particle density, $\eta$ the strength of interaction of $\eta/r^{\epsilon}$ -potential, and $D$ the spatial dimension of the system. The usual coherence decay in infinite systems has the form of ${\rm exp}(-t/\tau_{\rm d})$, which is used to define the decoherence time \cite{UZ89, zu91}. It appears that the decay profile of the form of $\exp \left(-(t/\tau_{\rm d} )^{ f(D,\epsilon)}\right)$ is more accurate than the usual one.

The starting point of simulations is a $D$-dimensional ``box'' whose volume is $l^{D}$. $N$ particles are placed randomly in this box. These particles are in fixed places, and they interact with each other according to the Hamiltonian (\ref{f0}). The size of the box is related to the number of particles $N$ and the particle density $n_{\rho}$ by
\begin{equation}
l=\left(\frac{N}{n_{\rho}}\right)^{1/D}.
\end{equation}
Without lack of generality and to ease numerics, we choose to study an initial state that contains complete superpositions, so $a_{k}=b_{k}=1/\sqrt{2}$. Thus, the closed and finite system has idealistic coherence $\Xi_{\rm id}=1$ and idealistic entropy $S_{\rm id}=0$. The interactions between particles have the form of $g_{lk}=\eta/r_{lk}^{\epsilon}$, where $r_{lk}$ is the distance between $l^{\rm th}$ and $k^{\rm th}$ particle. Inserting these conditions to Eq. (\ref{eq9}) results in the realistic coherence
\begin{equation}
\Xi_{\rm re}(t)=\frac{1}{N}\sum_{l=1}^N \prod_{k=1, \neq l}^N \cos \left(\frac{2\eta t}{r_{lk}^{\epsilon}}\right).
\end{equation}

Second, we use a least squares fit on $\Xi_{\rm re}(t)$ using a function 
\begin{equation}
\xi(t)=(1-c){\rm e}^{-(t/t_{\rm d})^{C}}+c
\label{eqfit}
\end{equation}
that allows fluctuations around the average level $c$ of $\Xi_{\rm re}(t)$. The average level $c$ is the expectation value of $\Xi_{\rm re}(t)$ in the interval $[t_{1},t_{2}]$, where $\tau_{\rm d}\ll t_{1}$ and $\tau_{\rm d}\ll t_{2}-t_{1}$:
\begin{equation}
c=\langle \Xi_{\rm re}\rangle=\frac{1}{t_{2}-t_{1}}\int^{t_{2}}_{t_{1}}\Xi_{\rm re}(t)dt.
\end{equation}
The sum of squares error $\chi$ for the function of Eq. (\ref{eqfit}) describes the accuracy of the fit. On the last phase of numerical analysis, it will be used to obtain relative weights for data points. The data of one run consists of fitting parameters $t_{\rm d}$ (represents the estimate of decoherence time), $C$ (exponent that characterises the coherence decay of $r^{-\epsilon}$-dependent potential in a certain dimension $D$), $c$ (the fluctuation level of coherence) and $\chi$. Poincar\'e recurrence times $T_{\rm P}$ are calculated directly from $g_{lk}$s.

The procedure is repeated $U$ times. For $U=100$ we get a good statistical estimate of the behaviour of the system with particular parameter values $(N_{i}, \epsilon_{i}, D_{i})$. The dependencies about $\eta$ and $n_{\rho}$ are included in the unit of ``simulation time'', i.e., $T=\eta n_{\rho}^{\epsilon/D}t$. The parameter space that is studied thoroughly is $N \in [20 \dots 100] \wedge D \in [1 \dots 4] \wedge \epsilon \in [1 \dots 2]$ and also some additional points and slices from greater $N$-values, most notably with $N=200$: $D \in [1 \dots 10] \wedge \epsilon \in [1 \dots 10]$.

\section{Results \label{results}}

\begin{figure}[htb]
\includegraphics[width=\textwidth]{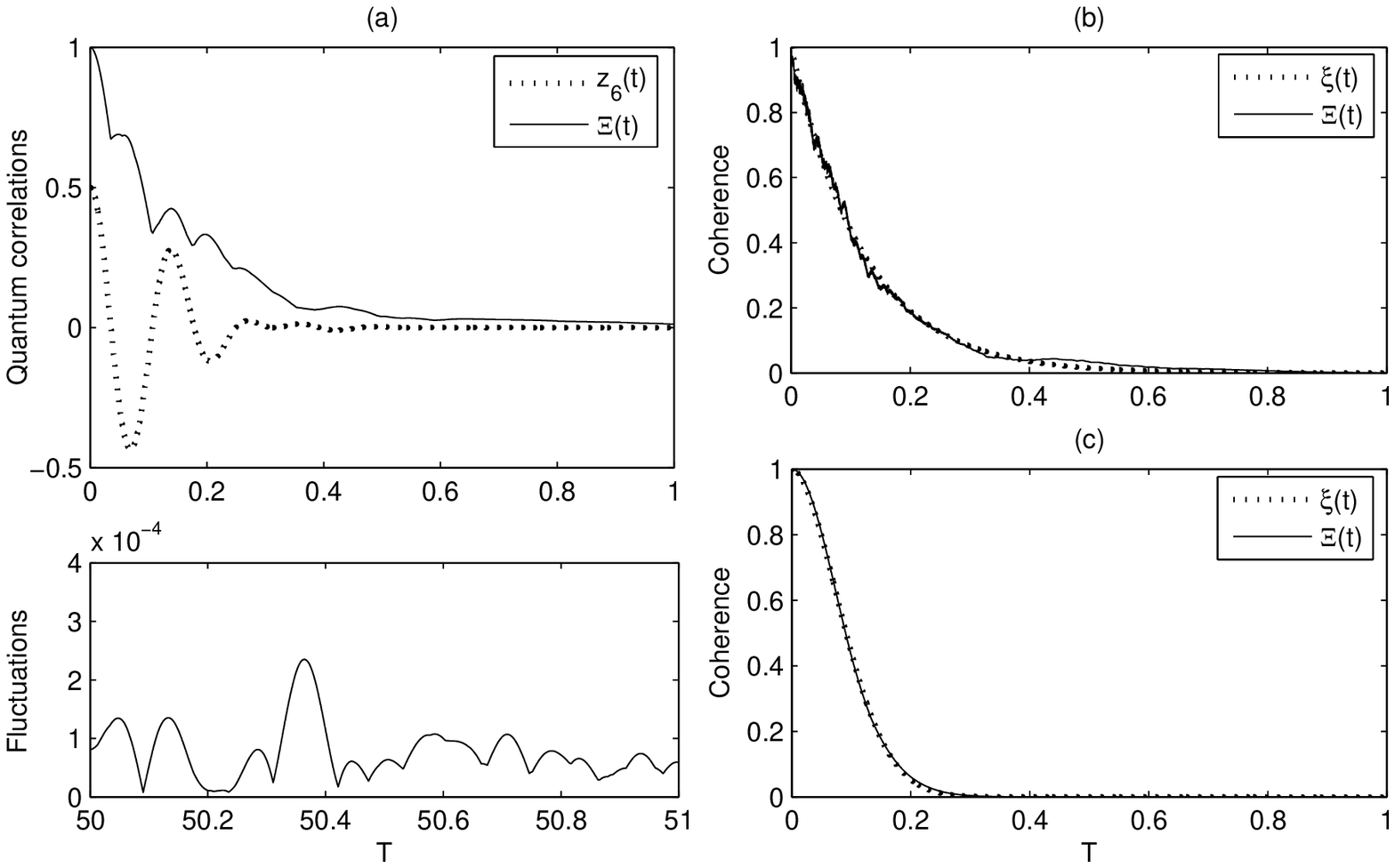}
\caption{The behaviour of coherence function $\Xi(t)$ in various scenarios (a) compared with single particle correlations $z_6(t)$ (of randomly chosen $6^{\rm th}$ particle) and long-time fluctuations with system parameters $N=20$, $D=1$, $\epsilon=1$, $n_{\rho}=1$, (b) compared with fitting function $\xi(t)$ with system parameters $N=100$, $D=1$, $\epsilon=1$, $n_{\rho}=1$, and (c) compared with fitting function $\xi(t)$ with system parameters $N=100$, $D=3$, $\epsilon=1$, $n_{\rho}=1$. The unit of simulation time is $T=\eta \left(2 n_{\rho} [{\rm m}] \right)^{\epsilon/D}t$, where $t$ is the real time. The fitting parameters (b) $t_{\rm d}=0.1212$, $C=1.0070$, $c=2.3\times 10^{-16}$, $\chi^2=0.8420$ and (c) $t_d=0.1114$, $C=1.8443$, $c=1.5\times 10^{-21}$, $\chi^2=0.0744$ demonstrate the accuracy of proposed coherence decay profile, and that fluctuations from theory are insignificant (c) when $D+1>\sqrt{N/200} \exp{(\epsilon/2)}$.}
\label{fig1}
\end{figure}

\begin{figure}[htb]
\includegraphics[width=\textwidth]{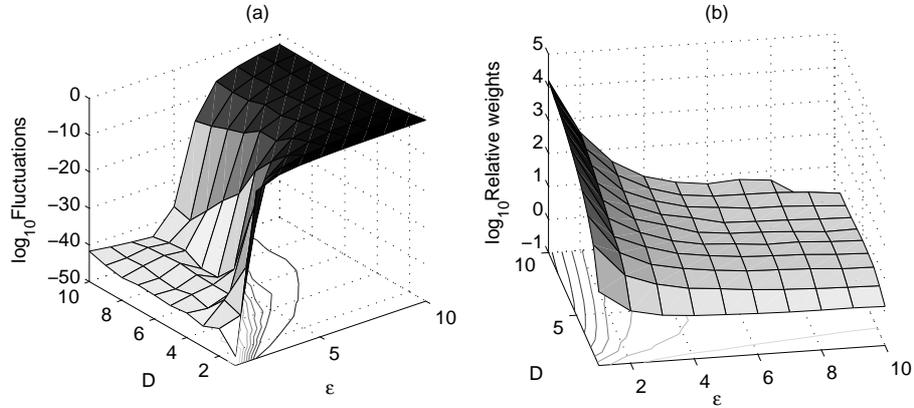}
\caption{Expectation value of fluctuations (a) and average of weights (b) in respect to dimension $D$ and the potential term $\epsilon$ with simulation parameter values $N=200$ and $n_{\rho}=1$. These plots demonstrate that the proposed coherence decay profile is very accurate especially if $D+1>\sqrt{N/200} \exp{(\epsilon/2)}$.}
\label{fig3}
\end{figure}

Numerical simulations clearly show that the proper coherence decay profile in our model is $\xi(t)=(1-c){\rm e}^{-(t/t_{\rm d})^{C}}+c$. Figure \ref{fig1} demonstrates the accuracy of the fit. The proposed decay profile $\xi(t)$ fits without significant fluctuations or deviations to simulation data if $D+1>\sqrt{N/200} \exp{(\epsilon/2)}$. The main factor for the observed feature is the behaviour of the distance measure of two coordinate points $l=\sqrt{\sum_i^D (x_i-x'_i)^2}$; it allows more particles within a fixed distance from the reference particle as the amount of spatial dimensions $D$ increases. Another factor is that as $\epsilon$ increases, the range of potential shortens and thus less particles within a fixed distance contribute to coherence decay. Figure \ref{fig3} presents the behaviour of expectation value of fluctuations and average of weights in respect to dimension $D$ and the potential term $\epsilon$. The minimum fluctuation level behaves as $\sim 10^{-N/5}$. Fluctuations begin to give a strong contribution if $D+1<\sqrt{N/200} \exp{(\epsilon/2)}$, so that fluctuation level with $D=1$ and $\epsilon=10$ is $\sim 0.3$ in which case fluctuations of coherence are quite dominant and thus that kind of system possesses considerable amount of its quantum behaviour despite the decoherence. As stated above, it is not surprising at all, since the short range of the potential isolates particles practically from all other particles except their nearest neighbours. However, the $\sim \sqrt{N}$ behaviour for the low-level fluctuation limit results in the interesting fact that even systems with $\epsilon=1$ -type of potentials do not lay on lowest coherence fluctuation level if the particle number is large enough. It seems that this result allows considerable coherence fluctuations for macroscopic objects, but the following facts should also be taken into account before claiming that these fluctuations will play significant role: (1) The floor of the fluctuation level drops as $\sim 10^{-N/5}$, so that the net effect of coherence fluctuations may still get smaller even if the system parameters do not lay on the diminishing bottom area. (2) The physical effect on the particular physical object is a sum of all interactions; including also those interactions that are on the bottom fluctuation level. 

The power $C$ of the exponential decay profile seems to have the upper limit $C\le 2$. As demonstrated in Figure \ref{fig1}, the usual coherence decay occurs if $D=\epsilon=1$. The functional forms of decoherence time $\tau_d$ and exponent $C$ are analysed in detail in Section \ref{dependence}. The sum of squared errors $\chi^2$ is used as the inverse of weight $w_i=\chi_i^{-2}$.

\subsection{About interactions and decoherence in different basis sets}

Let us study decoherence of the presented well-defined closed and finite model in a different basis set. So far, the study has considered the problem only in the $\sigma_z$-basis, but, the perfect superposition state in $\sigma_z$-basis is a well-defined eigenstate in $\sigma_x$-basis (spin up). According to the physical intuition, the results of nature should not depend on which way they are theoretically described. Thus, we should do a reality-check and study what happens to the density matrix in a general spin basis.

One can obtain any spin-basis via unitary transformation
\begin{equation}
U=\left(\begin{array}{cc}
\cos{\theta} & {\rm e}^{-i\phi}\sin{\theta} \\
{\rm e}^{i\phi}\sin{\theta} & -\cos{\theta}
\end{array}\right).
\end{equation}
The initial state of Eq. (\ref{f1}), with $a_k=b_k=\frac{1}{\sqrt{2}}$, is in the general spin-basis
\begin{equation}
\vert \Psi(0)\rangle_{\theta, \phi}=\prod^{N}_{k=1}\otimes \left( (\cos{\theta}+ {\rm e}^{-i\phi}\sin{\theta}) \vert+_{k}\rangle_{\theta, \phi} + (-\cos{\theta}+ {\rm e}^{i\phi}\sin{\theta}) \vert-_{k}\rangle_{\theta, \phi}\right).
\end{equation}
If one chooses $\theta=\pi/4$ and $\phi=0$, one gets $\sigma_x$-basis which is a product state of eigenstates with the probability of 1. This initial step of analysis illustrates some problems of the decoherent histories approach: how would a pure state with the probability of 1 experience decoherence? One can always find a basis set in which a superposition state appearing in another basis set is an eigenstate.

One should remember that decoherence is a {\it dynamical} process, and thus the interaction Hamiltonian is an equally important part of the problem as the initial state. In the present case, the Hamiltonian of Eq. (\ref{f0}) is in $\sigma_{\theta, \phi}$-basis
\begin{equation}
H=\hbar\sum_{j=1}^{N-1}\sum_{i=j+1}^{N}g_{ij}\sigma_{\theta,\phi}^j \otimes\sigma_{\theta, \phi}^i \prod_{k=1,\ne j,i}^N \otimes \mathbf{1}_k,
\end{equation}
where
\begin{equation}
\sigma_{\theta, \phi}=\left(\begin{array}{cc}
\cos^2{\theta}-\sin^2{\theta} &  {\rm e}^{-i\phi}\sin{2\theta} \\
{\rm e}^{i\phi}\sin{2\theta} & \sin^2{\theta}-\cos^2{\theta}
\end{array}\right).
\end{equation}
Thus, the reduced density matrix in a general spin-basis is
\begin{equation}
\rho(t)_{l, \theta, \phi}\!=\!\left(\!\begin{array}{cc}
\!\!\frac{1}{2}\!+\!\sin{\theta}\cos{\theta}\!\left({\rm e}^{-i\phi}z_l(t)\!+\!{\rm e}^{i\phi}z_l^*(t)\!\right)\!\!\!
& \!\!\!{\rm e}^{-i2\phi}\sin^2{\theta}z_l(t)\!-\!\cos^2{\theta}z_l^*(t)\!\! \\
\!\!{\rm e}^{i2\phi}\sin^2{\theta}z_l^*(t)\!-\!\cos^2{\theta}z_l(t)\!\!\! & \!\!\!\frac{1}{2}
\!-\!\sin{\theta}\cos{\theta}\!\left({\rm e}^{-i\phi}z_l(t)\!\!+\!{\rm e}^{i\phi}z_l^*(t)\!\right)\!\!
\end{array}\!\!\right).
\label{eqgen}
\end{equation}
Recall that in our example $z_l(t)=z_l^*(t)=\frac{1}{2}\prod_{k=1,\ne l}^N \cos{(2g_{lk}t)}$. The off-diagonal elements of the reduced density matrix (\ref{eqgen}) are proportional to $z_l(t)$ [or $z_l^*(t)$] in any basis that has off-diagonal elements, and therefore at least in the general spin-$\frac{1}{2}$ model the common idea of the relation between off-diagonal elements of the density matrix and quantum correlations (coherence) is valid, and the coherence function $\Xi(\rho_l,t)=2\vert z_l(t)\vert$ describes well the time evolution of coherence. As explained in Section \ref{cohdyn}, if $N\rightarrow \infty \wedge t\rightarrow \infty$, then $z(t),z^*(t)\rightarrow 0$, and thus the diagonalisation of the reduced density matrix (in infinite model) happens in all possible basis sets. In a finite model, Poincar\'e recurrences are present. The dynamical effects (decoherence and recoherence) do not depend on the chosen basis set, unlike in the decoherent histories approach, and thus, the preferred basis problem is not realised in density matrix formalism.

\subsection{Poincar\'e recurrence \label{recurrence}}

The interesting thing to notice is that our quantum system (of $N$ particles) is {\it a closed and finite quantum system}. Closedness results in the idealistic coherence $\Xi_{\rm id}$ being a constant of motion, and finiteness in that the realistic coherence $\Xi_{\rm re}(t)$ has a finite Poincar\'e recurrence time. There is an elementary procedure for an upper estimate of the recurrence time of a system that consists of $N$ subsystems with $M=0.5(N^2-N)$ fixed periods $T_{i}$. From the construct
\begin{equation}
\frac{T_{\rm unit}}{T_i}=\frac{n_i}{d_i},
\end{equation}
where $n_i$ and $d_i$ are smallest possible natural numbers, the upper limit of Poincar\'e recurrence time is obtained
\begin{equation}
T_{\rm P}=T_{\rm unit}\prod_i^M d_i.
\end{equation}

\begin{figure}[htb]
\includegraphics[width=\textwidth]{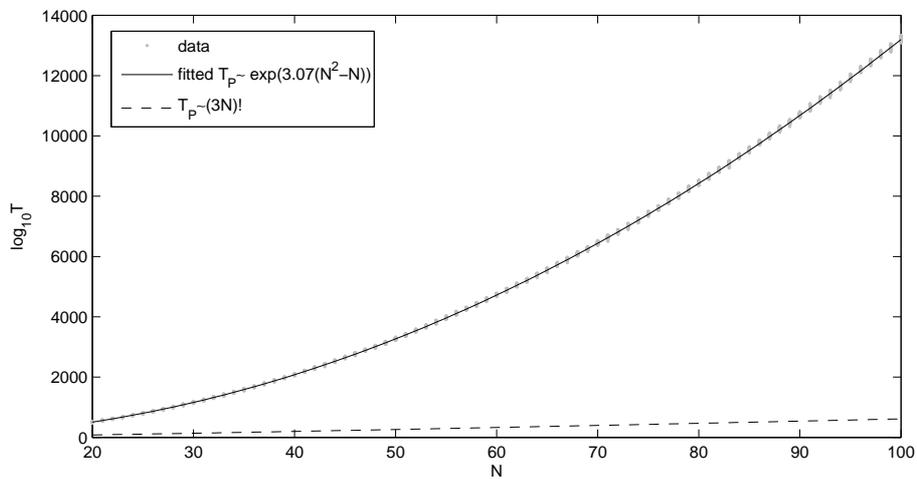}
\caption{An upper estimate of Poincar\'e recurrence times for simulation data (gray dots) with parameters $D=1$, $\epsilon=1$, and $n_{\rho}=1$. The unit of simulation time is $T=\eta \left(n_{\rho} [{\rm m}] \right)^{\epsilon/D}t$, where $t$ is the real time. Best fit (solid line) differs considerably from $\sim N!$ type of behaviour (dashed line).}
\label{fig2}
\end{figure}

The procedure of evaluating $T_{\rm P}$ of a particular simulation run is rather easy to do numerically, but with a drawback of limited accuracy. Thus, the presented Poincar\'e recurrence time analysis gives only an estimate of the magnitude of the recurrence times. The best fit for the data of the Figure \ref{fig2} is given by the function
\begin{equation}
T_{\rm P}(N,n_{\rho},\eta,\epsilon,D)=\pi\eta^{-1}n_{\rho}^{-\epsilon/D} {\rm e}^{3.07(N^2-N)}.
\end{equation}
In Ref. \cite{schlosshauer2004}, Poincar\'e recurrence time of a similar system is proportional to $T_{\rm P}\sim N!$. The difference between the estimates may be a result of limited numerical accuracy on applying our method. Another possibility is that there is unintentional ``double-counting''. Nevertheless, our estimate gives at least an upper estimate for recurrence time.

It seems clear that the system {\it may} return to its initial position, but the recurrence time grows fast with respect to $N$. For example, for $N=100$, $D=3$, $\epsilon=1$, $n_{\rho}=10^{30} ~{\rm m}^{-3}$ and $\eta=8.22 \times 10^{43} e^2 ~{\rm Hz\,m\,C^{-2}}$ (electro-magnetic interaction with charge of electron $e$) the estimated recurrence time is $T_{\rm P}\sim 10^{13183} ~{\rm s}$ and for $\sim N!$ the recurrence time is $T_{\rm P}\sim 10^{142} ~{\rm s}$ that is still a very long time compared with the age of the universe $T_{\rm U}\sim 4 \times 10^{17} ~{\rm s}$!

\subsection{The dependence of coherence decay on system variables \label{dependence}}

As shown before, the decay obeys well the coherence profile $\xi(t)=(1-c) {\rm e}^{-(t/t_{\rm d})^C}+c$. The floor of fluctuation level behaves as $c\sim 10^{N/5}$ if $D+1>\sqrt{N/200} \exp{(\epsilon/2)}$. To derive the dependence of coherence decay on relevant system variables has two separable steps remaining: to study power parameter $C$ and decoherence time parameter $t_{\rm d}$. 

\begin{figure}[htb]
\includegraphics[width=\textwidth]{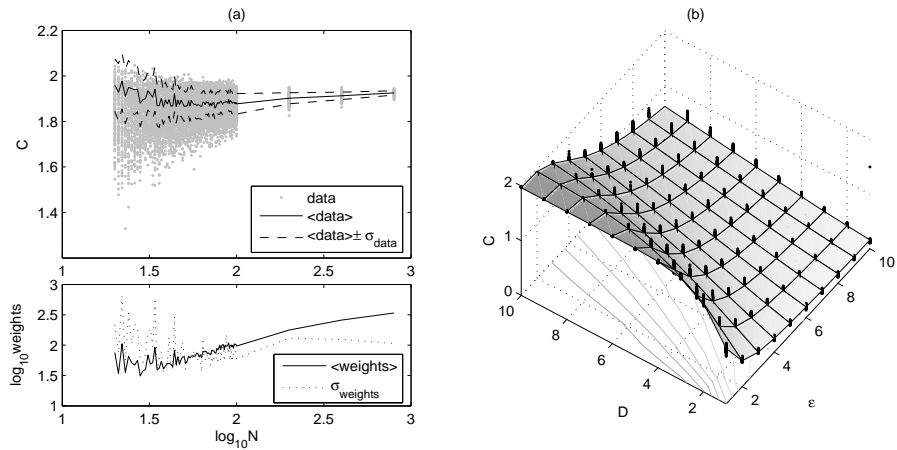}
\caption{The dependence of power parameter $C$ on system variables $N$, $\epsilon$ and $D$. The density is $n_{\rho}=1$. The slices of parameter space are $D=3$, $\epsilon=1$ (a), and $N=200$ (b). For certain parameter values ($N$, $D$, $\epsilon$) there exist 100 data points. It seems that there is no relation between $C$ and the number of particles $N$ (a). With small values of $N$, the standard deviation of weights is greater than average of weights, which explains deviations of weighted average from a constant value. With respect to dimension $D$ and potential parameter $\epsilon$, the best fit for power parameter $C$ is given by the function $f(D,\epsilon) =1.97\left(1-0.93\exp(-0.65D^{1.35} \epsilon^{-1.68})\right)$ (b). Note that for large $D$ and $\epsilon$, the deviation in data points gets wider.}
\label{fig4}
\end{figure}

The power parameter $C$ does not seem to be a function of the particle number $N$, as Figure \ref{fig4} clearly shows. The function
\begin{equation}
f(D,\epsilon) =1.97\left(1-0.93\exp{(-0.65D^{1.35}\epsilon^{-1.68})}\right)
\end{equation}
gives the best fit for the values of $C$ of the fitting function $\xi(t)$. As both $D$ and $\epsilon$ increase, the deviation of data points in parameter $C$ increases. 

Quite a good approximation of the functional behaviour of the decoherence time-related parameter $t_{\rm d}$ of $\xi(t)$ is given by
\begin{eqnarray}
\tau_{\rm d} \!\! & \!\!=\!\! & \!\! \eta^{-1}n_{\rho}^{-\epsilon/D}\left(\frac{N}{200}\right)^{-0.085(D-1)/\epsilon} \\
\!\!& \!\!\times\!\! & \!\!\!\! \left(0.29{\rm e}^{-0.79(D-1)\epsilon^{1/4} -0.13(\epsilon-3.4)^2} \!\!+ 0.17\frac{\epsilon\!-\!1}{2^\epsilon}((D\!-\!2)^2)^{0.19(\epsilon\!-\!1)} \!+\! \frac{0.03}{\epsilon^{1/2}} \!+\! 0.07\right). \nonumber
\label{dekoaika}
\end{eqnarray}
We assume that the accuracy of the fit $\tau_{\rm d}$ is limited to the close environment of the parameter space where $D,\epsilon \le 10$, since the form of the fit is very complex and most likely it does not follow the physical behaviour of the system accurately enough; it is only a fit on the observed behaviour. Moreover, the deviation in data points increases as both $D$ and $\epsilon$ increase. The most evident reason for this is the fact that $N=200$ particles is still a small number in a short-range interaction parameter space ($\epsilon>3$) when the number of particles pro dimension is very small ($N/D<50$). In the future, this inconvenience should be easily overcome as the available computational capacity increases. The behaviour of $\tau_{\rm d}$ with respect to the particle number $N$ is more reliable because of the few data points with $N=400$ and $N=800$, and because particle number-related functions factorise out of the complex form in equation (\ref{dekoaika}).

Despite the fact that for each parameter point there are only 100 simulation runs and there are accuracy problems as both $D$ and $\epsilon$ increase, the behaviour of $f(D,\epsilon)$ and $\tau_{\rm d}(N,n_{\rho},\eta,\epsilon,D)$ is quite reliable for physically relevant cases. Physically most interesting results are:
\begin{itemize}
\item The decoherence time $\tau_{\rm d}$ behaves as $\sim N^{-{\rm const.}(D-1)/\epsilon}$ with respect to the particle number $N$. Is there a constant level that decoherence time approaches as the particle number increases? To answer this question, one seems to need simulations with $N$ that is at least one or two orders of magnitude greater.

Moreover, the fit has $(D-1)$. Is this ``$-1$'' a physical property of the model or a result of numerics? We suspect the latter, but there was no suitable fit of $\tau_{\rm d}$ without ``$-1$''.

\item The upper limit of the exponent $f(D,\epsilon)$ of the decay profile $\Xi(t)=\exp[-(t/\tau_{\rm d})^{f(D,\epsilon)}]$ is $\sim 2$.

\item The decoherence time $\tau_{\rm d}$ seems to have a lower limit with respect to $D$ and $\epsilon$. For $N=200$, it is around $\sim 0.07$ in units of simulation time $\eta \left(n_{\rho} [{\rm m}] \right)^{\epsilon/D}t$.

\item The behaviour of $\tau_{\rm d}$ in the parameter space $(D,\epsilon)$ is dominated by two different trends. One trend is a Gaussian peaking near $(D=1,\epsilon=3.4)$ and the other one is a runaway solution when both $D$ and $\epsilon$ are large. We suspect that at least a part of that Gaussian is a physical property of the model, but also that most of the runaway solution is caused by too few particles in the numerics.

\item For $D=3$, $N=100$ and $n_{\rho}=10^{30} ~{\rm m}^{-3}$, the electro-magnetic interaction $(\epsilon=1)$ results in decoherence time $\tau_{\rm d}=4.3\times 10^{-17} ~{\rm s}$ and exponent $f(D, \epsilon)=1.87$. For spin-spin interaction $(\epsilon=3)$ and $\eta=3.27\times 10^{-26} ~{\rm m^3 \,Hz}$ (for $^6{\rm Li}$) with the same setup, the decoherence time is $\tau_{\rm d}=3.3\times 10^{-5} ~{\rm s}$ with exponent $f(D,\epsilon)=0.80$. For pure spin-spin interaction, the experimental observation of coherence decay seems possible in few-particle systems, but only iff particles are isolated enough from other (stronger) interactions, most notably from electro-magnetic interaction.

\end{itemize}

While a more sophisticated study needs more computational power, obvious results of this research are that our definition of coherence is confirmed by the case-study, and that the coherence decay profile is $\Xi_{\rm re}(t)=(1-c)\exp{\left[-(t/\tau_{\rm d})^f\right]}+c$ in the Heisenberg model with $\eta/ r^{\epsilon}$ -dependent interactions in various dimensions.

\section{Discussion \label{discussion}}

In the past, there have been arguments that while decoherence in closed systems is practically a working concept, closed systems do not, in principle, experience decoherence. A famous argument against decoherence in closed systems is from Bell \cite{om94, be75}. Bell's argument is mainly that if the universe starts in a pure state, it will always remain in a pure state, no matter how quickly the off-diagonal elements of the reduced density matrix decrease and how small they will become. Bell claims that this gives, {\it in principle}, a possibility to make such a measurement that will show quantum interference. However, there is not such a measurement device, even in principle, that can perform the measurement on the universe. We demonstrated this with the Clever Chinchilla, who is a supernatural-like observer outside the universe (closed system). She can guess right the wave function of the universe, along with the interaction Hamiltonian, and then calculates the time evolution of the universe. In quantum measurements, one often mixes the view of the Clever Chinchilla and the Real Observer who studies readings of a measurement apparatus inside the universe. The Clever Chinchilla is not correlated with the universe, while our Real Observer is totally correlated with it, and thus their views about a particular quantum event may differ. The Real Observer can never achieve the Clever Chinchilla's knowledge about the universe. He has no means to solve by measurements what is the whole wave function of the universe. He only observes projections (possibly very complicated ones) of the wave function of the universe. For a detailed discussion about Bell's argument, one is encouraged to study references \cite{om94, be75}.

However, the universe is a closed system and decoherence is an observed phenomenon in our universe. Thus, (de)coherence should be uniquely quantifiable. In this paper, we present a unique way to quantify coherence. The growth of entropy has the same dynamical origin as decoherence. In conclusion, they both are equally correct descriptions, also in closed systems. Moreover, they grasp different aspects of a quantum system. Coherence as defined in this article expands knowledge of the properties of quantum systems beyond what is possible to obtain from entropy. First of all, coarse-grained coherence $\Xi_{\rm re}$ could be used to calculate the decoherence time of the system. It is impossible to obtain decoherence time  from entropy. Second, it is the quantity that restricts the observation of quantum properties of a system more dramatically than entropy. They both give different kinds of constraints that a closed quantum system must obey. It is possible to have two systems with equal coherence but different entropy, and it is also possible to have two systems with equal entropy but different coherence. The latter group of systems is very interesting for quantum mechanical engineering -- especially for studies related to quantum computation. Third, coherence may be useful in quantifying entanglement. Quantum computing still lacks a unique method to quantify ``the usefulness of entanglement'' beyond two entangled systems. Our proposal for mutual entanglement $E=\Xi_{\rm id}-\Xi_{\rm re}$ may be worth studying in this respect.

The view, as the Real Observer inside the universe perceives it, can be obtained by using realistic coherence and entropy, which are effective theory descriptions about the knowledge of the Clever Chinchilla. Thus, the use of reduced density matrices is justified. Decoherence seems to work also {\it in principle} in closed quantum systems, and consistency problems can be avoided by studying decoherence of reduced density matrices instead of decoherent histories.

To demonstrate the usefulness of coherence as defined in this article, we solved the time evolution of (realistic) coherence in the closed and finite spin model and noticed that the decay profile of coherence does not necessarily follow the commonly expected exponential decay. The trivial result that the diagonalisation of reduced density matrices occurs in all possible basis sets suggests that in ``real'' quantum experiments and measurements, observed noise or ``measurement errors'' in the behaviour of highly quantum correlated systems may be results of a weak, unmodelled interaction that causes decoherence. Increased understanding on this topic may lead to better modelled quantum measurements, better measurement technology and technical solutions which may help to keep coherent systems as coherent as possible in the real universe.

\section*{Acknowledgements}
The author would like to thank Kalle-Antti Suominen for guiding him into the scientific world, and Magnus Ehrnrooth Foundation, the Academy of Finland (project 206108), and the Graduate School for Modern Optics and Photonics for funding. He would like to give special thanks to Heikki Lindroos, whose expertise in computer hardware made the heavy finite model simulations possible and who also provided a considerable amount of the processor time used for the simulations. He also thanks John Calsamiglia for questions that have inspired him to try to find answers, if there are any, Matt Mackie for constructive criticism, and Riina Kosunen and Anna Dannenberg for helping to understand how to apply terminological methods in theoretical physics.

\pagebreak
\section*{Figures}

\begin{figure}[htb]
\includegraphics[width=\textwidth]{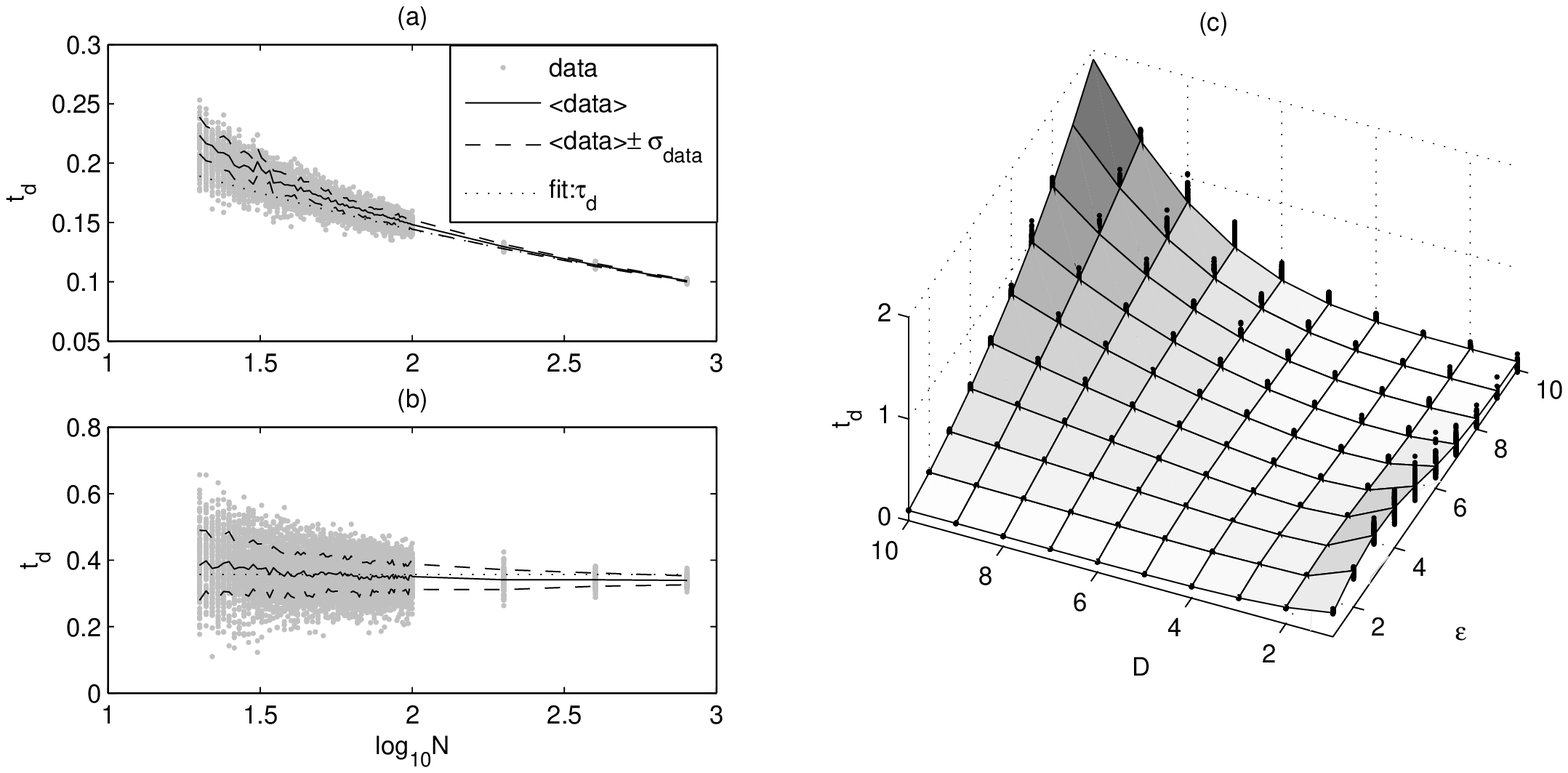}
\caption{The dependence of decoherence time parameter $t_{\rm d}$ on system variables $N$, $\epsilon$ and $D$. The slices of parameter space are $D=3$, $\epsilon=1$ (a), $D=1$, $\epsilon=2$ (b), and $N=200$ (c). The density is $n_{\rho}=1$ and the unit of simulation time is $[t_{\rm d}]=\eta \left(n_{\rho} [{\rm m}] \right)^{\epsilon/D}t$, where $t$ is the real time. For certain parameter values ($N$, $D$, $\epsilon$) there exist 100 data points. A rather good approximation of functional behaviour of $t_{\rm d}$ is given by $\tau_{\rm d}$ of equation (\ref{dekoaika}). Note that for small $N$ (a,b), large $D$ and $\epsilon$ (c), the deviation in data points gets wider.}
\label{fig5}
\end{figure}


\begin{thebibliography}{} %

   \bibitem{sc35}
    E. Schr\"odinger,
    Die Naturwissenschaften {\bf 23}, 807-812, 823-828, 844-849 (1935).

   \bibitem{cl83}
    A. O. Caldeira and A. J. Leggett,
    Physica A {\bf 121}, 587 (1983);
    Phys. Rev. A {\bf 31}, 1059 (1985).

   \bibitem{UZ89}
    W. G. Unruh and W. H. Zurek,
    Phys. Rev. D {\bf 40}, 1071 (1989).

   \bibitem{zu91}
    W. H. Zurek,
    Phys. Today {\bf 44}(10), 36 (1991).

   \bibitem{hpz92}
    B. L. Hu, J. P. Paz, and Y. Zhang,
    Phys. Rev. D {\textbf{45}}, 2843 (1992).

   \bibitem{alzp95}
    J. R. Anglin, R. Laflamme, W. H. Zurek, and J. P. Paz,
    Phys. Rev. D {\bf 52}, 2221 (1995).

   \bibitem{dm2006}
    O. Dannenberg and M. Mackie,
    Phys. Rev. A {\bf 74}, 053601 (2006).

   \bibitem{schlosshauer2004}
    M. Schlosshauer,
    Rev. Mod. Phys. {\bf 76}, 1267 (2004).

   \bibitem{sc2005}
    M. Schlosshauer,
    Phys. Rev. A {\bf 72}, 012109 (2005).

   \bibitem{br96}
    M. Brune, E. Hagley, J. Dreyer, X. Ma\^{i}tre, A. Maali, C. Wunderlich, J. M. Raimond, and S. Haroche,
    Phys. Rev. Lett. {\bf 77}, 4887 (1996).

   \bibitem{nemes}
    M. C. Nemes,
    Introduction: Experimental and Theoretical Status of Decoherence,
    in: Decoherence and Entropy in Complex Systems, edited by H.-T. Elze 
    (Springer-Verlag, Heidelberg, 2004), and references therein.

   \bibitem{dh92}
    H. F. Dowker and J. J. Halliwell,
    Phys. Rev. D {\bf 46}, 1580 (1992).

   \bibitem{gmh93}
    M. Gell-Mann and J. B. Hartle,
    Phys. Rev. D {\bf 47}, 3345 (1993).

   \bibitem{bh1996}
    T. A. Brun and J. J. Halliwell,
    Phys. Rev. D {\bf 54}, 2899 (1996).

   \bibitem{bh99}
    T. A. Brun and J. B. Hartle,
    Phys. Rev. D {\bf 60}, 123503 (1999).

   \bibitem{halliwell}
    J. J. Halliwell,
    Some Recent Developments in the Decoherent Histories Approach to Quantum Theory,
    in: Decoherence and Entropy in Complex Systems, edited by H.-T. Elze
    (Springer-Verlag, Heidelberg, 2004).

  \bibitem{cl1997}
    M. Castagnino and R. Laura,
    Phys. Rev. A {\bf 56}, 108 (1997);
    Int. J. Theor. Phys. {\bf 39}, 1737 (2000);
    Phys. Rev. A {\bf 62}, 022107 (2000).

   \bibitem{lc1998}
    R. Laura and M. Castagnino,
    Phys. Rev. E {\bf 57}, 3948 (1998);
    Phys. Rev. A {\bf 57}, 4140 (1998).

   \bibitem{castagnino1999}
    M. Castagnino,
    Int. J. Theor. Phys. {\bf 38}, 1333 (1999).

   \bibitem{lcb1999}
    R. Laura, M. Castagnino, and R. I. Betan,
    Physica A {\bf 271}, 357 (1999).

   \bibitem{cl2005}
    M. Castagnino and O. Lombardi,
    Int. J. Theor. Phys. {\bf 42}, 1281 (2003);
    Stud. Hist. Philos. Mod. Phys. {\bf 35}, 73 (2004);
    Phys. Rev. A {\bf 72}, 012102 (2005).

   \bibitem{zu82}
    W. H. Zurek,
    Phys. Rev. D {\bf 26}, 1862 (1982).

   \bibitem{om94}
    R. Omn\`es,
    The Interpretation of Quantum Mechanics
    (Princeton University Press, Princeton, 1994).

   \bibitem{goldstein1998}
    S. Goldstein, 
    Phys. Today {\bf 51}(3), 42, {\bf 51}(4), 38 (1998).

   \bibitem{dk96}
    H. F. Dowker and A. Kent,
    J. Stat. Phys. {\bf 82}, 1575 (1996).

   \bibitem{k96}
    A. Kent,
    Phys. Rev. A {\bf 54}, 4670 (1996).

   \bibitem{k97}
    A. Kent,
    Phys. Rev. Lett. {\bf 78}, 2874 (1997).

   \bibitem{ghirardi1999}
    A. Bassi and G. C. Ghirardi,
    Phys. Lett. A {\bf 257}, 247 (1999).

   \bibitem{barrett2003}
    J. A. Barrett,
    Everett's Relative-State Formulation of Quantum Mechanics, 
    in: The Stanford Encyclopedia of Philosophy, edited by E. N. Zalta (Spring 2003 Edition),
    http://plato.stanford.edu/archives/spr2003/entries/qm-everett/.

   \bibitem{rs1975}
    M. Reed and B. Simon,
    Fourier Analysis, Self-Adjointness
    (Academic Press, New York, 1975).

   \bibitem{gi60}
    J. W. Gibbs,
    Elementary Principles in Statistical Mechanics
    (Dover, New York, 1960).

   \bibitem{sk93}
    L. Sklar,
    Physics and Chance: Philosophical Issues in the Foundations of Statistical Mechanics
    (Cambridge University Press, Cambridge, 1993).

   \bibitem{mcgrawhill}
    McGraw-Hill Dictionary of Scientific and Technical Terms, 6th Edition
    (AccessScience@McGraw-Hill, http://www.accessscience.com, November 12 2006).

   \bibitem{ze}
    J.-W. Pan, D. Bouwmeester, M. Daniell, H. Weinfurter, and A. Zeilinger,
    Nature {\bf 403}, 515-519 (2000).

   \bibitem{zu81}
    W. H. Zurek,
    Phys. Rev. D {\bf 24}, 1516 (1981).

   \bibitem{alav4}
    Noether's theorem states that invariance of the Lagrangian under a group of 
    continuous transformations implies the conservation of some quantity. See 
    our Ref. \cite{no18}.

   \bibitem{no18}
    E. Noether,
    Invariante Variationsprobleme
    Nachr. d. K\"onig. Gesellsch. d. Wiss. zu G\"ottingen, Math-phys. Klasse, 235-257 (1918);
    English translation in 
    M. A. Travel, 
    Transport Theory and Statistical Physics {\bf 1}, 183-207 (1971).

   \bibitem{gag1997}
    F. H. Gaioli, E. T. G. Alvarez, and J. Guevara,
    Int. J. Theor. Phys. {\bf 36}, 2167-2207 (1997).

   \bibitem{gs1995}
    B. M. Garraway and K.-A. Suominen,
    Rep. Progr. Phys. {\bf 58}, 365-419 (1995).

   \bibitem{ms1991}
    P. Meystre and M. Sargent III,
    Elements of Quantum Optics, 2nd edition
    (Springer-Verlag, Heidelberg, 1991).

   \bibitem{wm1994}
    D. F. Walls and G. J. Milburn,
    Quantum Optics
    (Springer-Verlag, Heidelberg, 1994).

   \bibitem{carmichael1999}
    H. J. Carmichael,
    Statistical Methods in Quantum Optics 1: Master Equations and Fokker-Planck Equations
    (Springer-Verlag, Heidelberg, 1999).

   \bibitem{bp2002}
    H.-B. Breuer and F. Petruccione,
    The Theory of Open Quantum Systems
    (Oxford University Press, Oxford, 2002).

   \bibitem{bekenstein1981}
    J. D. Bekenstein,
    Phys. Rev. D {\bf 23}, 287-298 (1981);
    Phys. Rev. Lett. {\bf 46}, 623-626 (1981);
    Phys. Rev. D {\bf 30}, 1669-1679 (1984).

   \bibitem{lloyd2002}
    S. Lloyd,
    Phys. Rev. Lett. {\bf 88}, 237901 (2002).

   \bibitem{ee59}
    P. Ehrenfest and T. Ehrenfest,
    The Conceptual Foundations of the Statistical Approach in Mechanics
    (Cornell University Press, Ithaca NY, 1959).

   \bibitem{sk74}
    L. Sklar,
    Space, Time and Spacetime
    (University of California Press, California, 1974).

   \bibitem{pe79}
    R. Peierls,
    Surprises in Theoretical Physics
    (Princeton University Press, Princeton, 1979).

   \bibitem{re95}
    M. Redhead,
    From Physics to Metaphysics
    (Cambridge University Press, Cambridge, 1995).

   \bibitem{br03}
    S. Brush,
    Kinetic Theory of Gases
    (Imperial College Press, London, 2003).

   \bibitem{bo96}
    L. Boltzmann,
    Entgegnung auf die w\"armetheoretischen Betrachtungen des Hrn. E. Zermelo,
    Annalen der Physik und Chemie {\bf 57}, 773-784 (1896);
    reprinted and translated in \cite{br03}, 393-402.

   \bibitem{alav2}
    More elementary details about information, entropy and quantum physics can 
    be found in almost all quantum information textbooks, e.g., in Ref. \cite{ss05}.

   \bibitem{ss05}
    S. Stenholm and K.-A. Suominen,
    Quantum Approach to Informatics
    (John Wiley \& Sons, Hoboken NJ, 2005).

   \bibitem{go31}
    K. G\"odel, 
    \"Uber formal unentscheidbare S\"atze der Principia Mathematica und 
    verwandter Systeme, I, 
    Monatshefte f\"ur Mathematik und Physik, {\bf 38}, 173-198 (1931);
    English translation in 
    J. van Heijenoort, 
    From Frege to G\"odel: A Source Book in Mathematical Logic, 1879-1931
    (Harvard University Press, Harvard, 2002).

   \bibitem{da03}
    O. Dannenberg, R. Kosunen, and A. Dannenberg
    (unpublished).

   \bibitem{entropiasta}
    However, the concept of {\it non-extensive entropy} has been studied in statistical physics and 
    information theory. For more details, see e.g., Refs. \cite{jizba, tsallis} and references therein.

   \bibitem{jizba}
    P. Jizba,
    Information Theory and Generalized Statistics,
    in: Decoherence and Entropy in Complex Systems, edited by H.-T. Elze
    (Springer-Verlag, Heidelberg, 2004).

   \bibitem{tsallis}
    Y. S. Weinstein, C. Tsallis, and S. Lloyd,
    On the Emergence of Nonextensivity at the Edge of Quantum Chaos,
    in: Decoherence and Entropy in Complex Systems, edited by H.-T. Elze
    (Springer-Verlag, Heidelberg, 2004).

   \bibitem{alav}
    In their model the particles that form an environment do not interact with each other.

   \bibitem{be75}
    J. S. Bell,
    Helv. Phys. Acta {\bf 48}, 93 (1975).

\end{thebibliography}
\end{document}